\documentclass[twoside,twocolumn,english,prb,a4paper]{revtex4}
\usepackage[T1]{fontenc}
\usepackage[latin1]{inputenc}
\usepackage{geometry}
\usepackage{babel}
\usepackage{graphics}

\makeatletter

\providecommand{\LyX}{L\kern-.1667em\lower.25em\hbox{Y}\kern-.125emX\@}

\makeatother
\begin{document}

\title{The importance of friction in the description of low-temperature
dephasing}

\author{by Florian Marquardt}

\email{Florian.Marquardt@unibas.ch}

\affiliation{Department of Physics and Astronomy, University of Basel, Klingelbergstrasse
82, CH-4056 Basel, Switzerland}

\begin{abstract}
We discuss the importance of the real part of the Feynman-Vernon influence
action for the analysis of dephasing and decay near the ground state
of a system which is coupled to a bath. Using exactly solvable linear
quantum dissipative systems, it is shown how the effects of the real
and the imaginary part (describing friction and fluctuations, respectively)
may cancel beyond lowest-order perturbation theory. The resulting
picture is extended to a qualitative discussion of nonlinear systems
and dephasing of degenerate fermions. We explain why dephasing rates
will, in general, come out finite at zero temperature if they are
deduced from the imaginary part of the action alone, a procedure which
is reliable only for highly excited states.
\end{abstract}

\date{July 30th, 2002}

\maketitle

\section{Introduction}

The Feynman-Vernon influence functional\cite{feynvern,feynhibbs}
plays a prominent role in discussions of dephasing that aim to go
beyond a simple master equation approach. In principle, the influence
functional takes into account every effect of the environment (the
{}``bath'') on the system under consideration. This includes heating,
dephasing, friction and renormalization effects (changing the external
potential or the effective mass of a particle). Its popularity arises
not only from the fact that it constitutes an exact approach, but
also from the direct physical meaning which it acquires in some situations.
In particular, this holds for typical interference experiments, where
two wave packets describing a single particle follow two different
trajectories in order to be recombined later on. For such a case,
the influence functional is equal to the overlap between the two different
bath states which result due to the particle moving along either one
of the semiclassical trajectories. If the bath can distinguish between
the two paths, it acts as a which-way detector, and the diminished
magnitude of the overlap directly gives the ensuing decrease of the
visibility of the interference pattern. Writing the overlap in the
form of an exponential, \( \exp (iS_{R}-S_{I}) \), one can obviously
conclude that only the imaginary part \( S_{I} \) of the {}``influence
action'' is responsible for this decrease and, therefore, \( S_{I} \)
alone describes dephasing \emph{in such a situation}. Furthermore,
if the system is subject to a fluctuating external \emph{classical}
field, only \( S_{I} \) remains, while \( S_{R} \) vanishes identically.
This is because \( S_{I} \) is due to the fluctuations of the bath,
while \( S_{R} \) stems from the back-action of the bath onto the
system (including friction effects), which is absent for an external
noise field. The approach of evaluating \( S_{I} \) along semiclassical
trajectories has been successfully applied to the calculation of dephasing
rates in many situations\cite{AAK,sai,cohen}. 

However, near the ground state of a system, an analysis along these
lines is likely to fail. Qualitatively, this may already be deduced
from the fact that dropping \( S_{R} \) means replacing the environment
by an artificial fluctuating classical field whose correlation function
includes the zero-point fluctuations of the original quantum bath.
Therefore, even at zero temperature, this fluctuating field will,
in general, heat the system, which directly leads to dephasing. The
role of \( S_{R} \) is to counteract this effect. In this article,
it is our aim to display explicitly, in a detailed manner, the necessity
of keeping the real part \( S_{R} \) of the influence action in discussions
of dephasing and decay near the ground state of a system.

This issue derives its importance partly from the fact that the evaluation
of \( S_{I} \) along semiclassical trajectories has proven to be
an efficient way of extracting dephasing rates in the problem of weak
localization\cite{AAK,sai} in the limit of high temperatures, when
zero-point fluctuations of the bath may be omitted. In contrast, at
low temperatures, a single-particle semiclassical calculation may
become invalid, since it neglects the Pauli principle which is known
to play an important role for the inelastic scattering of electrons
and is not included in the Feynman-Vernon influence functional. However,
recently an extension of the influence functional to the case of a
many-fermion system has been derived, using an exact procedure\cite{GZ,GZ_CL,GZ_PB}
and including the Pauli principle. This permitted a discussion of
dephasing in a disordered metal even for the case of low temperatures.
Following the general strategy of earlier works\cite{chakravarty}
dealing with the high-temperature case, the dephasing rate was deduced
from \( S_{I} \) in a semiclassical calculation and found to be finite
even at zero temperature. Since the {}``orthodox'' theory\cite{AAK}
had predicted a vanishing rate in the limit \( T\rightarrow 0 \),
the new results prompted considerable criticism\cite{AAGcritique,VavAmbeg,cohenimry,BelitzKirckpatrick},
which mostly emphasized technical aspects of impurity averaging or
used perturbation theory to arrive at different conclusions. In the
present work, we want to clarify some essential aspects of the roles
of \( S_{R} \) and \( S_{I} \), using physically much more transparent
exactly solvable models.

We want to emphasize that {}``zero-temperature dephasing'' as such
is perfectly possible: If one prepares a system in any superposition
of excited states and couples it to a bath, it will, in general, decay
towards its ground state by spontaneous emission of energy into the
bath (at \( T=0 \)). This will destroy any coherent superposition,
thus leading to dephasing. Considerations of this kind are particularly
relevant for quantum-information processing, where one necessarily
deals with nonequilibrium situations involving excited qubit states
of finite energy. The situation is different for the weak-localization
problem (and similar transport interference effects): There, one is
interested in the zero-frequency limit of the system's linear response,
which depends on the coherence properties of arbitrarily low-lying
excited states. It is the subtleties associated with a path-integral
description of these situations which we want to address in this work.

The article is organized as follows: After a brief review of the influence
functional and its meaning in semiclassical situations (sec.~\ref{section2}),
we will rewrite the expressions for \( S_{I} \) and the Golden Rule
decay rate, in order to compare the two (sec.~\ref{section3}). In
doing so, we will closely follow the analysis of Cohen and Imry\cite{cohen,cohenimry}.
Then, we will show how and why the effects of \( S_{R} \) and \( S_{I} \)
may compensate each other in the derivation of a decay rate starting
from the path integral expression for the time-evolution of the density
matrix (sec.~\ref{section4}), even though they cannot cancel each
other in the influence action. There, it may be observed that drawing
conclusions about decay directly from the exponent \( iS_{R}-S_{I} \)
of the influence functional is usually not possible. The crucial cancellation
takes place at a later stage of the calculation, after proper integration
over the fluctuations around the classical paths and after averaging
over the initial state. In order to prove that this compensation takes
place not only in lowest order perturbation theory, we will then specialize
to exactly solvable linear dissipative systems, i.e. the damped harmonic
oscillator (sec.~\ref{linearmodels}) and the free particle (sec.~\ref{qubrownmotion}).
We will also point out that there is an important difference between
the oscillator and the free particle: Dropping \( S_{R} \) has a
much more drastic effect on the former, leading to an artificial finite
decay rate of the ground state at zero temperature. However, in order
to discuss the importance of \( S_{R} \) for the calculation of dephasing
rates (involving decay of excited states), we have to extend our analysis
to nonlinear models (sec. \ref{section7}). We will explain in a more
qualitative fashion why the essential insights gained from the exactly
solvable models should remain valid both for the nonlinear models
as well as for systems of degenerate fermions.

\section{The influence functional and dephasing in simple situations}

\label{section2}We are interested in the time-evolution of the reduced
density matrix of a system with coordinate \( q \) which is coupled
to some environment (the bath). If system and bath had been uncoupled
prior to \( t=0 \), then the density matrix at a later time \( t \)
is linearly related to that at time \( 0 \): 

\begin{equation}
\label{rhoJ}
\rho (q^{>}_{t},q^{<}_{t},t)=\int dq_{0}^{>}dq_{0}^{<}\, J(q_{t}^{>},q_{t}^{<}|q_{0}^{>},q_{0}^{<};t)\rho (q_{0}^{>},q_{0}^{<},0)\, .
\end{equation}

The propagator \( J \) on the right-hand-side of this equation is
given by 

\begin{equation}
\label{prop}
J=\int \! \! \! \int Dq^{>}Dq^{<}\exp \left[ i(S_{0}^{>}-S_{0}^{<})\right] \exp \left[ iS_{R}-S_{I}\right] \, .
\end{equation}

Here we have set \( \hbar \equiv 1 \) and introduced an abbreviated
notation: The path integral extends over all {}``forward'' paths
\( q^{>}(\cdot ) \) running from the given value of \( q^{>}_{0} \)
to \( q^{>}_{t} \), likewise for the {}``backward'' paths \( q^{<}(\cdot ) \).
The value of the action of the uncoupled system, evaluated along \( q^{>(<)}(\cdot ), \)
is denoted by \( S_{0}^{>(<)} \). The second exponential in Eq.~(\ref{prop})
is the Feynman-Vernon influence functional\cite{feynvern,feynhibbs}.
Both \( S_{R} \) and \( S_{I} \) are real-valued functionals that
depend on both paths simultaneously. The influence functional is the
overlap of bath states which have time-evolved out of the initial
bath state \( \chi _{0} \) under the action of either \( q^{>}(\cdot ) \)
or \( q^{<}(\cdot ) \):

\begin{equation}
e^{iS_{R}-S_{I}}\equiv \left\langle \chi [q^{<}(\cdot )]|\chi [q^{>}(\cdot )]\right\rangle \, .
\end{equation}

At zero temperature, the state \( \chi _{0} \) is the ground state
of the unperturbed bath, while at finite temperatures an additional
thermal average over \( \chi _{0} \) has to be performed. From this
representation, it follows\cite{feynvern,feynhibbs} that \( S_{R} \)
changes sign on interchanging \( q^{>} \) and \( q^{<} \) while
\( S_{I} \) always remains nonnegative - the magnitude of the overlap
can only be decreased compared with its initial value of one.

The meaning of \( S_{R} \) and \( S_{I} \) becomes particularly
transparent for an interference setup, where two wave packets \( \Psi _{>} \)
and \( \Psi _{<} \) travel along two different paths \( q_{cl}^{>}(\cdot ) \)
and \( q^{<}_{cl}(\cdot ) \). We want to assume a situation which
can be described semiclassically, i.e. the wave-length is supposed
to be much smaller than the size of a wave packet and this again is
much smaller than the typical separation between the paths and the
typical dimensions over which the external potential changes. Then,
it suffices to evaluate \( S_{R} \) and \( S_{I} \) just for the
combination of these two paths, since the fluctuations around them
are comparatively unimportant. The environment will affect the interference
pattern on the screen mainly by changing the interference term\cite{sai}: 

\begin{eqnarray}
\rho (x,x,t) & \approx  & \left| \Psi _{>}(x,t)\right| ^{2}+\left| \Psi _{<}(x,t)\right| ^{2}+\nonumber \\
 &  & 2\, Re\left[ \Psi _{>}(x,t)\Psi ^{*}_{<}(x,t)e^{iS_{R}-S_{I}}\right] \, .\label{simpleinterference} 
\end{eqnarray}

In this equation, \( \Psi _{>(<)}(x,t) \) are assumed to represent
the unperturbed time-evolution of the wave packets. To a first approximation,
\( S_{R,I} \) do not enter the {}``classical'' terms in the first
line of Eq. (\ref{simpleinterference}), since \( S_{R,I}[q_{cl}^{>},q^{>}_{cl}]=0 \).
Deviations from this first approximation stem from the integration
over fluctuations away from \( q^{>(<)}_{cl} \) and describe, for
example, mass- and potential renormalization as well as slowing-down
of the wave packet due to friction. Obviously, \( S_{I}\equiv S_{I}[q^{>}_{cl},q^{<}_{cl}] \)
determines directly the decrease in visibility of the interference
pattern while \( S_{R}\equiv S_{R}[q^{>}_{cl},q^{<}_{cl}] \) only
gives a phase-shift. Averaging the interference pattern over different
configurations of the external potential (appropriate for impurity
averaging in mesoscopic samples) may further decrease the visibility.
However, if the two paths are time-reversed copies of each other (as
is the case in discussions of weak-localization\cite{AAK,sai}), the
phase difference between \( \Psi _{>} \) and \( \Psi _{<} \) vanishes
in a time-reversal invariant situation, so that the impurity-average
over the corresponding phase factor does not lead to a suppression
in the situation without the bath. On the other hand, the average
over \( \exp (iS_{R}) \) will, in general, decrease further the magnitude
of the interference term\cite{notesravg}. In any case, given the
simple physical picture presented here, one would not expect \( S_{I} \)
and \( S_{R} \) to be able to cancel each other's effects, since
they represent, respectively, the real and imaginary part of an exponent.

For the special case of a force \( \hat{F} \) deriving from a bath
of harmonic oscillators (linear bath), with vanishing average \( \left\langle \hat{F}\right\rangle =0 \)
and a linear interaction \( \hat{V}=-\hat{q}\hat{F} \), we have:

\begin{eqnarray}
S_{I} & = & \int _{0}^{t}\! \! \! dt_{1}\! \! \! \int _{0}^{t_{1}}\! \! \! dt_{2}\left( q^{>}_{1}-q^{<}_{1}\right) Re\left\langle \hat{F}_{1}\hat{F}_{2}\right\rangle \left( q^{>}_{2}-q^{<}_{2}\right) \label{silin} \\
S_{R} & = & -\! \int _{0}^{t}\! \! \! dt_{1}\! \! \! \int _{0}^{t_{1}}\! \! \! dt_{2}\left( q^{>}_{1}-q^{<}_{1}\right) Im\left\langle \hat{F}_{1}\hat{F}_{2}\right\rangle \left( q^{>}_{2}+q^{<}_{2}\right) \label{srsi} 
\end{eqnarray}

Here we have used the notation \( q^{>}_{1}\equiv q^{>}(t_{1}) \).
The angular brackets denote averaging over the equilibrium state of
the unperturbed bath. \( S_{I} \) only depends on the symmetrized
part of the bath correlator \( Re\left\langle \hat{F}(\tau )\hat{F}(0)\right\rangle =\left\langle \left\{ \hat{F}(\tau ),\hat{F}(0)\right\} \right\rangle /2 \),
which becomes the classical correlator \( \left\langle F(\tau )F(0)\right\rangle  \)
for the case of classical Gaussian random noise. In the latter case,
\( S_{R} \) vanishes and the influence functional \( \exp (-S_{I}) \)
is simply the classical average of the phase factor 

\begin{equation}
\exp \left[ i\int _{0}^{t}(q^{<}(\tau )-q^{>}(\tau ))F(\tau )d\tau \right] \, .
\end{equation}

If one drops \( S_{R} \) in a dephasing calculation (noting that
it does not enter dephasing in semiclassical situations such as those
that can be described by Eq.~(\ref{simpleinterference})), one effectively
replaces the quantum bath by a classical fluctuating force whose correlator
is determined by the symmetrized part of the quantum correlator. This
contains the zero-point fluctuations, since \begin{eqnarray}
\left\langle \left\{ \hat{Q}(\tau ),\hat{Q}(0)\right\} \right\rangle \propto  &  & \nonumber \\
\left( 2n(\omega )+1\right) \cos (\omega \tau )=\coth (\omega /2T)\cos (\omega \tau ) &  & 
\end{eqnarray}
 for the coordinate \( \hat{Q} \) of a single bath oscillator of
frequency \( \omega  \), with \( n(\omega ) \) being the Bose distribution
function that vanishes at \( T=0 \).

Although replacing the quantum bath by a classical noise force seems
to be a drastic step, it can lead to correct predictions for the dephasing
rate in semiclassical situations such as the one discussed above,
even at zero temperature. It is instructive to observe how this comes
about in an exactly solvable model. A particularly simple situation
is the one analyzed by Caldeira and Leggett\cite{clwavepackets} (see
also Ref.~\onlinecite{lossmullen}). They considered a damped quantum
harmonic oscillator, where the initial state consisted of a superposition
of Gaussian wavepackets, one centered at the origin, the other at
a distance \( z \). In the course of time, the displaced wavepacket
oscillates back and forth in the oscillator potential well. Whenever
the packets overlap, an interference pattern results (due to the difference
in the respective momenta). The environment leads both to damping
and dephasing, where the latter typically proceeds at a much faster
rate. For our purposes, we are interested in the limit of small damping,
where, to a first approximation, the center-of-mass motion of the
wave packets is not appreciably altered by friction in the period
of the oscillation. In this case, it turns out that, indeed, the result
predicted by the approximation Eq.~(\ref{simpleinterference}) for
the attenuation of the interference pattern is correct. This can be
seen in Ref.~\onlinecite{clwavepackets} by taking the limit of weak
damping (\( \gamma \rightarrow 0,\, \omega \rightarrow \omega _{R} \))
in their exact result for the exponent of the attenuation factor (see
Eq.~(2.13) of Ref.~\onlinecite{clwavepackets}) and comparing this
to \( S_{I} \) (see Eq.~(\ref{silin})), which is to be evaluated
for the pair of classical paths followed by the two wave packets,
\( q_{cl}^{>}(t)=z\, \cos (\omega t) \) and \( q_{cl}^{<}(t)\equiv 0 \).
In terms of the quantity \( C_{00} \) listed in appendix \ref{appc00}
of the present work, both results may be obtained by multiplying \( C_{00} \)
by \( \sin (\tilde{\omega }t)^{2} \) and setting \( \gamma =0,\, \tilde{\omega }=\omega  \).
In addition, in order to obtain \( S_{I}[q^{>}_{cl},q^{<}_{cl}] \),
the factors \( \sin (\omega (t-t_{1})) \) and \( \sin (\omega (t-t_{2})) \)
must be replaced by \( \cos (\omega t_{1}) \) and \( \cos (\omega t_{2}) \).
The results obviously coincide for the points in time when the wave-packets
meet (\( \omega t=\pi /2+n\pi  \)). We emphasize that the correspondence
to the semiclassical result holds only for a situation far away from
the ground state of the system, where at least one of the wave packets
is in a superposition of highly excited oscillator states. 

At zero bath temperature, the dephasing is purely due to spontaneous
emission, as pointed out in the discussion of Ref.~\onlinecite{clwavepackets}.
Energy is transferred from the oscillating wave-packet into the bath
and the system relaxes towards lower-energy states. The dephasing
rate is proportional to the total rate \( \Gamma _{out} \) at which
the system leaves a given energy level\cite{clwavepackets}. Of course,
this rate may be obtained using the Golden Rule (i.e. a master equation
description) only for weak coupling, but the qualitative picture seems
to be general\cite{clwavepackets}. In the correct description of
the physical situation considered here, the rate \( \Gamma _{out} \)
is given entirely by the rate of spontaneous emission, \( \Gamma ^{em}_{sp} \).
However, if \( S_{R} \) is neglected, then \( S_{I} \) describes
a classical noise force (equivalent to the zero-point fluctuations
of the bath at \( T=0 \)) and there will be both induced emission
and absorption, proceeding at equal rates \( \Gamma ^{em}_{ind}=\Gamma ^{abs}_{ind} \).
Therefore, the system is also excited by the bath in that approximation.
Nevertheless, we have pointed out above that the total dephasing rate
comes out right. The reason is the following: The rate \( \Gamma _{out} \)
is the same in both cases, because the rate of spontaneous emission
in the correct description is exactly twice that of induced emission
in the approximation:

\begin{equation}
\Gamma _{out}=\Gamma ^{em}_{sp}\equiv \Gamma ^{em}_{ind}+\Gamma ^{abs}_{ind}\, .
\end{equation}
At this point, it is easy to see the physical reason why such an approximation
may fail near the ground state of the system. Then, the transitions
downwards in energy may be blocked, which completely suppresses \( \Gamma ^{em}_{sp} \),
but not \( \Gamma ^{abs}_{ind} \). We will make this argument more
precise in the following sections.

\section{Decay rates from \protect\( S_{I}\protect \) and Golden rule: Dependence
on the spectra of bath and system motion}

\label{section3}The growth of \( S_{I} \) with time depends on the
spectral density of the bath fluctuations at frequencies which appear
in the system's motion. Formally, this can be seen by introducing
the spectrum of the system motion related to the given pair of paths
\( q^{>(<)} \) (following the ideas of Refs.~\onlinecite{cohen,cohenimry}):

\begin{eqnarray}
P(\omega ,s) & \equiv  & \frac{1}{2\pi }\int _{-\infty }^{+\infty }d\tau \, e^{i\omega \tau }(q^{>}(s+\frac{\tau }{2})-q^{<}(s+\frac{\tau }{2}))\nonumber \\
 &  & \times (q^{>}(s-\frac{\tau }{2})-q^{<}(s-\frac{\tau }{2}))\label{systemspec} 
\end{eqnarray}

Here, \( s\equiv (t_{1}+t_{2})/2 \) and \( \tau \equiv t_{1}-t_{2} \)
are sum and difference times. Therefore, \( P \) is the Fourier transform
with respect to \( \tau  \) of the \( q \)-dependent terms in \( S_{I} \)
(Eq. (\ref{silin})). We take \( q^{>(<)}(t') \) to be zero whenever
\( t' \) falls outside the range \( [0,t] \). 

Furthermore, we define \( \left\langle FF\right\rangle _{\omega } \)
to be the Fourier-transform of the symmetrized correlator of \( \hat{F} \), 

\begin{equation}
\left\langle FF\right\rangle _{\omega }\equiv \frac{1}{4\pi }\int _{-\infty }^{+\infty }d\tau \, e^{i\omega \tau }\left\langle \left\{ \hat{F}(\tau ),\hat{F}(0)\right\} \right\rangle \, ,
\end{equation}

so it is real and symmetric in frequency. The same holds for the system
spectrum \( P(\omega ,s) \). 

Using this, \( S_{I} \) can be expressed in the following way: 

\begin{equation}
\label{overlapspectra}
S_{I}=\pi \int _{0}^{t}ds\int _{-\infty }^{+\infty }d\omega \, \left\langle FF\right\rangle _{\omega }P(-\omega ,s)\, .
\end{equation}

If the dependence of \( P \) on \( s \) is not essential, then \( S_{I} \)
grows linearly with time \( t \), at a rate given by the {}``overlap
of bath and system spectra''. Similar expressions can be derived
for a spatially inhomogeneous fluctuating force\cite{cohen,cohenimry},
leading to an additional \( k \)-dependence. 

Applying this kind of reasoning to a \emph{single} electron moving
in a dirty metal\cite{cohen}, the system motion is found to contain
frequencies up to (at least) \( 1/\tau _{el} \), such that the growth
of \( S_{I} \) with time depends on the bath-spectrum up to this
cutoff, including the zero-point fluctuations of the corresponding
high-frequency bath modes (which become important at low temperatures).
As explained above (sec. \ref{section2}), the description of dephasing
in terms of \( S_{I} \) alone may be trusted in a semiclassical situation,
where the electron is in a highly excited state. This holds even at
zero temperature, when the qualitative physical picture is essentially
the same as in the model of oscillating wave packets due to Caldeira
and Leggett, discussed above: Dephasing is due to spontaneous emission
of energy into the bath. The contribution of frequencies up to \( 1/\tau _{el} \)
then implies that the spontaneous emission (and the resulting dephasing)
is facilitated by the impurity scattering. The physics behind this
is well-known in another context: In quantum electrodynamics, the
emitted radiation would be called {}``bremsstrahlung'', since it
is the scattering off an external potential that induces the electron
to emit radiation.

On the other hand, one can express in a similar manner decay rates
from a simple Golden Rule calculation\cite{cohenimry}. For the decay
of an initial state \( \left| i\right\rangle  \) , we have

\begin{equation}
\label{decayrate}
\Gamma _{i}=2\pi \int d\omega \, \left\langle \hat{F}\hat{F}\right\rangle _{\omega }\left\langle \hat{q}\hat{q}\right\rangle ^{(i)}_{-\omega }\, .
\end{equation}

Here \( \left\langle \hat{F}\hat{F}\right\rangle _{\omega } \) and
\( \left\langle \hat{q}\hat{q}\right\rangle ^{(i)}_{-\omega } \)
are the Fourier transforms of the \emph{unsymmetrized} correlators
of \( \hat{F} \) and \( \hat{q} \), taken in the equilibrium state
of the bath or the initial state of the system, respectively:

\begin{eqnarray}
\left\langle \hat{F}\hat{F}\right\rangle _{\omega } & \equiv  & \frac{1}{2\pi }\int d\tau \, e^{i\omega \tau }\left\langle \hat{F}(\tau )\hat{F}(0)\right\rangle \nonumber \\
\left\langle \hat{q}\hat{q}\right\rangle ^{(i)}_{\omega } & \equiv  & \frac{1}{2\pi }\int d\tau \, e^{i\omega \tau }\left\langle i\right| \hat{q}(\tau )\hat{q}(0)\left| i\right\rangle 
\end{eqnarray}

The important point to notice is that, near the ground state of the
system, the quantum correlator \( \left\langle \hat{q}\hat{q}\right\rangle _{\omega }^{(i)} \)
is very asymmetric in frequency space, as the system can mostly be
excited only (\( \omega >0 \)). At low temperatures, the same holds
for the bath. Thus, the decay rate (\ref{decayrate}), containing
the product of correlators evaluated at \( \omega  \) and \( -\omega  \),
is very much suppressed below the value that it would acquire if either
the bath-correlator or the system-correlator were symmetrized (becoming
symmetric both in time and frequency). Since dropping \( S_{R} \)
is equivalent to symmetrizing the bath correlator and, furthermore,
semiclassical calculations give a symmetric spectrum \( P(\omega ,s) \)
of the system motion as well, it becomes clear why there are situations
when the decay rate, as deduced from \( S_{I} \), is finite at low
temperatures, while the Golden-rule decay rate vanishes. The question
then arises whether any procedure that amounts to symmetrization of
the correlators, thus leading to drastically wrong results for Golden
Rule decay rates at zero temperature, may be justified to discuss
dephasing using a path-integral approach. Observations such as this
have led Cohen and Imry to conclude that, in their semiclassical analysis
of electron dephasing inside a metal\cite{cohenimry}, the contribution
of the bath's zero-point fluctuations to the dephasing rate should
be dropped. Their argument was not a mathematical proof, but rather
drawn from physical intuition and analogies. In the following sections,
we try to elucidate the importance of \( S_{R} \) in descriptions
of dephasing and decay near the ground state of a system, demonstrating
exactly how a cancellation between \( S_{R} \) and \( S_{I} \) may
arise.

\section{Cancellation of \protect\( S_{R}\protect \) and \protect\( S_{I}\protect \)
in lowest-order perturbation theory}

\label{section4}We consider a system which is in its ground state
at \( t=0 \) before being coupled to a linear bath. From the Golden
Rule, we expect there to be no finite decay-rate for the ground state
at zero-temperature. It is well-known\cite{feynhibbs} how to derive
a master equation from the influence functional, by expanding it to
lowest order in the exponent, \( iS_{R}-S_{I} \). We display that
calculation here in order to point out where and how \( S_{R} \)
and \( S_{I} \) do cancel. The time-evolution of the probability
to find the system in its unperturbed ground state is:

\begin{eqnarray}
 &  & P_{0}(t)\equiv \left\langle \Psi _{0}\right| \hat{\rho }(t)\left| \Psi _{0}\right\rangle \nonumber \\
 & = & \int dq_{t}^{>}dq_{t}^{<}\rho _{0}(q^{<}_{t},q^{>}_{t})\rho (q_{t}^{>},q_{t}^{<},t)\nonumber \\
 & = & \int Dq^{>}Dq^{<}\rho _{0}(q^{<}_{t},q^{>}_{t})e^{i(S_{0}^{>}-S_{0}^{<})+iS_{R}-S_{I}}\rho _{0}(q_{0}^{>},q_{0}^{<})\nonumber \\
 & \approx  & 1+\int Dq^{>}Dq^{<}e^{i(S^{>}_{0}-S^{<}_{0})}\rho _{0}(q^{<}_{t},q^{>}_{t})\nonumber \\
 &  & \, \times (iS_{R}-S_{I})\rho _{0}(q_{0}^{>},q_{0}^{<})\, .\label{eq13} 
\end{eqnarray}

Here \( \rho _{0} \) is the density matrix of the unperturbed ground
state, and the path integration also includes integrals over the initial
and final coordinates. Now one can insert the expressions for \( S_{R} \)
and \( S_{I} \) from Eq.~(\ref{srsi}). The integration over the
trajectories and the endpoints produces the (\emph{unsymmetrized})
correlators of the system coordinate, such as (for \( t_{1}>t_{2} \)):

\begin{eqnarray}
\int Dq^{>}e^{iS_{0}^{>}}\Psi ^{*}_{0}(q_{t}^{>})q^{>}(t_{1})q^{>}(t_{2})\Psi _{0}(q_{0}^{>})= &  & \nonumber \\
\left\langle \Psi _{0}\right| \hat{q}(t_{1})\hat{q}(t_{2})\left| \Psi _{0}\right\rangle e^{-iE_{0}t}\, . &  & 
\end{eqnarray}

Evidently, both the unperturbed action \( S_{0} \) and the integrations
involving the state \( \Psi _{0} \) are essential for obtaining the
correct correlator. In the long-time limit, we obtain the decay rate
which also follows from the Golden Rule (\ref{decayrate}): \begin{widetext}

\begin{equation}
\label{pathmaster}
P_{0}(t)\approx 1-\int _{0}^{t}dt_{1}\int _{0}^{t_{1}}dt_{2}\, 2\, Re\left[ \left\langle \hat{F}(t_{1})\hat{F}(t_{2})\right\rangle \left\langle \Psi _{0}\left| \hat{q}(t_{1})\hat{q}(t_{2})\right| \Psi _{0}\right\rangle \right] \approx 1-\Gamma _{0}t\, .
\end{equation}

\end{widetext}Only the term growing linearly in time has been retained
at the end of this equation. 

Note that, of course, there will always be a small reduction in the
probability of finding the unperturbed ground state after switching
on the interaction, since the ground state of the coupled system contains
contributions from other system states as well. This point has been
discussed in more detail recently for the case of the damped harmonic
oscillator\cite{buettHO}. It does not contribute to \( \Gamma _{0} \)
which describes the decay linear in time. 

We emphasize that the decay of \( P_{0}(t) \) is not governed by
\( S_{I} \) alone: If \( S_{R} \) were dropped, then only the (real-valued)
symmetrized version of the bath correlator would appear in Eq.~(\ref{pathmaster}).
In that case, the decay rate would only depend on the (real-valued)
symmetric part of the system correlator as well. Therefore, a finite
decay rate of the ground state would result even at zero temperature,
where really there would have been none. The physical reason is the
heating introduced by the classical noise field, whose correlator
contains the zero-point fluctuations of the original quantum bath,
as we have discussed before. 

On a formal level, we may argue that the decay of the density matrix
is not governed by \( S_{I} \) alone since the {}``weighting factor''
that is used when {}``averaging'' over many paths contains the \emph{phase}
factor \( \exp (i(S_{0}^{>}-S_{0}^{<})) \). Therefore, such an average
is not the same as a classical average, for which the decay could
not be overestimated by dropping \( S_{R} \), since then we could
use \( \left| \left\langle \exp (iS_{R}-S_{I})\right\rangle \right| \leq \left\langle \exp \left( -S_{I}\right) \right\rangle  \).
Furthermore, the integration over the density matrix of the initial
state is obviously essential, as it is needed to produce the correct
form of the system correlator. Both facts mean that it would be premature
to draw any conclusions about dephasing and decay at an early stage
of the calculation, by merely looking at the influence action.

\section{Exactly solvable linear systems: {}``Cancellation to all orders''
for the damped oscillator}

\label{linearmodels}In order to show how \( S_{R} \) and \( S_{I} \)
can cancel also beyond lowest-order perturbation theory, we will now
turn to linear quantum dissipative systems. Although the exact solutions
for the damped harmonic oscillator and the free particle have been
well-known for a long time\cite{callegg}, we will review the essential
steps in the derivation, pointing out where \( S_{R} \) and \( S_{I} \)
do enter. We will first turn to the oscillator, for which the Golden
Rule result suggests that one would obtain an artificial decay of
the ground state at zero temperature if \( S_{R} \) were neglected.
For simplicity, our discussion is restricted to \( T=0 \), since
that is the limit where these effects show up most clearly.

For any linear system which is linearly coupled to a bath of oscillators,
the propagator \( J \) for the density matrix (Eq.~(\ref{prop}))
can be evaluated exactly, since the integration over system paths
is Gaussian. In fact, \( J \) is found to be given by the {}``semiclassical''
result, i.e. an exponential containing the action along stationary
paths, multiplied by a prefactor which does not contain the endpoints
of the paths (a specialty of linear systems):

\begin{equation}
\label{Jsemi}
J(q^{>}_{t},q^{<}_{t}|q^{>}_{0},q^{<}_{0};t)=\frac{1}{N(t)}e^{iS[q^{>}_{cl},q^{<}_{cl}]}\, ,
\end{equation}

with \( S=S_{0}^{>}+S_{0}^{<}+S_{R}+iS_{I} \). The paths \( q^{>(<)}_{cl} \)
make the full action stationary,

\begin{equation}
\label{stat}
\frac{\delta S}{\delta q^{>}_{cl}}=\frac{\delta S}{\delta q^{<}_{cl}}=0\, ,
\end{equation}

and fulfill boundary conditions of the form \( q^{>}_{cl}(0)=q^{>}_{0} \).
The prefactor \( N(t) \) can be obtained most easily from the condition
for normalization of the density matrix, 

\begin{equation}
\label{normal}
\int dq_{t}\, J(q_{t},q_{t}|q_{0},q_{0};t)=1\, .
\end{equation}

\( N(t) \) is found to be independent of \( S_{I} \), but it does
depend on \( S_{R} \) (see, for example, the general proof in app.
E of Ref.~\onlinecite{GZ_CL}). 

In the following, we will turn to the special case of the \emph{Ohmic}
bath, which leads to a velocity-proportional friction force and has
a power spectrum rising linearly at low frequencies (for \( T=0 \)):

\begin{equation}
\label{ohmbathdef}
\left\langle FF\right\rangle _{\omega }=\frac{\eta |\omega |}{2\pi }\, .
\end{equation}

We will argue below that, in the case of the damped harmonic oscillator,
no essential qualitative result will be changed by using other bath
spectra, unless these have an excitation gap exceeding the oscillator
frequency.

As usual, we introduce the center-of-mass coordinate \( R(\tau )\equiv (q^{>}(\tau )+q^{<}(\tau ))/2 \)
and the difference coordinate \( r(\tau )\equiv q^{>}(\tau )-q^{<}(\tau ) \)
in order to write down the equations obtained from (\ref{stat}),
for the case of the Ohmic bath: 

\begin{eqnarray}
\frac{d^{2}r}{d\tau ^{2}}-\gamma \frac{dr}{d\tau }+\omega _{0}^{2}r & = & 0\label{req} \\
\frac{d^{2}R}{d\tau ^{2}}+\gamma \frac{dR}{d\tau }+\omega _{0}^{2}R & = & \! \! \! -i\! \! \! \int _{0}^{t}\! \! \! d\tau '\, Re\! \left\langle \hat{F}(\tau )\hat{F}(\tau ')\right\rangle \! r(\tau ')\label{Req} 
\end{eqnarray}

Here \( \gamma \equiv \eta /m \) is the damping rate and \( \omega _{0} \)
is the unperturbed frequency of the oscillator. Note that the second
equation leads to a complex-valued solution \( R(\cdot ) \). It is
also possible to formulate the calculation slightly differently\cite{callegg,hakimambeg},
by using stationary solutions with respect to the real part of \( S \)
only. In any case, inserting the solutions \( R \) (\( \equiv R_{cl} \))
and \( r \) (\( \equiv r_{cl} \)) into the action \( S \) shows
that the imaginary part of the action \( S_{cl}\equiv S[q_{cl}^{>},q^{<}_{cl}] \)
is determined directly only by \( S_{I} \). As expected, \( S_{cl} \)
turns out to be a bilinear expression in the endpoints \( R_{t},\, R_{0},\, r_{t},\, r_{0} \):

\begin{eqnarray}
Re\, S_{cl} & = & S_{0}^{>}+S_{0}^{<}+S_{R}=\nonumber \label{res} \\
 &  & R_{t}r_{t}L_{tt}+R_{t}r_{0}L_{t0}+R_{0}r_{t}L_{0t}+R_{0}r_{0}L_{00}\label{res} \\
Im\, S_{cl} & = & S_{I}=r_{0}^{2}C_{00}+2r_{0}r_{t}C_{0t}+r_{t}^{2}C_{tt}\, .\label{ims} 
\end{eqnarray}

All the coefficients are real-valued functions of time. For our purposes,
it is sufficient to know that the entries of the matrix \( L \) are
independent of \( S_{I} \) (while they do depend on the damping \( \gamma  \)
from \( S_{R} \)). In contrast, the entries of \( C \) depend both
on \( S_{I} \) and \( S_{R} \). The dependence on \( S_{R} \) arises
only because the path \( r \), which is inserted into Eq.~(\ref{silin}),
is affected by the friction described by \( S_{R}, \) see Eq.~(\ref{req}).
The entries of \( C \) are proportional to the strength of the bath
fluctuations (i.e. the magnitude of \( \left\langle FF\right\rangle _{\omega } \)).
Explicit expressions for all of these coefficients\cite{5} can be
found in Ref.~\onlinecite{callegg} (cf. their section 6). For reference,
the quantities used in the following discussion have been listed in
appendix \ref{appc00}.

At this point, evaluation of the coefficients for \( Im\, S_{cl} \)
shows that they will grow in time beyond all bounds, regardless of
whether the path \( r \) is calculated by taking into account \( S_{R} \)
or neglecting it (setting \( \gamma =0 \) in Eq.~(\ref{req})).
However, no conclusion about dephasing and decay can be drawn from
this, since the foregoing discussions lead us to expect that any potential
cancellation between \( S_{R} \) and \( S_{I} \) will take place
only \emph{after} proper integration over the initial density matrix.
This can be deduced from the derivation of the Golden Rule decay rate
given in the preceding section. It is also very reasonable that \( S_{I} \)
grows without bounds for any semiclassical path \( r(\cdot ) \),
since all paths contribute more or less to the time-evolution of all
eigenstates, and decay of excited states is perfectly correct in the
model of the damped harmonic oscillator.

The initial ground state density matrix has the form 

\begin{eqnarray}
\rho _{0}(q_{0}^{>},q_{0}^{<})\propto \exp \left( -(q_{0}^{>2}+q_{0}^{<2})/(4\left\langle \hat{q}^{2}\right\rangle _{0})\right)  &  & \nonumber \\
=\exp \left( -(R_{0}^{2}+r_{0}^{2}/4)/(2\left\langle \hat{q}^{2}\right\rangle _{0})\right) \, , &  & \label{indens} 
\end{eqnarray}

with \( \left\langle \hat{q}^{2}\right\rangle _{0}=1/(2m\omega _{0}) \)
as the square of the ground state width. After performing the Gaussian
integrals over \( R_{0} \) and \( r_{0} \), which are needed to
evaluate Eq.~(\ref{rhoJ}), we obtain the final result for the density
matrix \( \rho (q^{>}_{t},q^{<}_{t},t) \) at a time \( t \) after
switching on the interaction with the bath. It has the general form

\begin{equation}
\frac{1}{\tilde{N}(t)}\exp \left( -(aR_{t}^{2}+bR_{t}r_{t}+cr_{t}^{2})\right) \, .
\end{equation}

In particular, the prefactor of \( R_{t}^{2} \) gives the width of
the probability distribution at time \( t \),

\begin{equation}
\left\langle \hat{q}^{2}(t)\right\rangle =\left( 2a\right) ^{-1}\, .
\end{equation}

Let us now look at the behaviour of the width in order to analyze
the effect of dropping \( S_{R} \) in the calculation. The expression
for \( \left\langle \hat{q}^{2}(t)\right\rangle  \) has been derived
already by Caldeira and Leggett\cite{callegg} (cf. their Eq.~6.34):

\begin{equation}
\label{width}
\left\langle \hat{q}^{2}(t)\right\rangle =\left( 2C_{00}+m\omega _{0}/2+L_{00}^{2}/(2m\omega _{0})\right) /L_{t0}^{2}\, .
\end{equation}

This expression clearly contains quantities both from the imaginary
part (\( C_{00} \)) and from the real part (\( L_{00},\, L_{t0} \))
of the action. Formally speaking, real and imaginary parts have become
intermixed when integrating over \( R_{0} \), due to quadratic completion
of the expression \( iR_{0}(r_{t}L_{0t}+r_{0}L_{00})-m\omega _{0}R_{0}^{2} \)
in the exponent (the former term is from Eq. (\ref{res}), while the
latter term is from the initial density matrix, Eq. (\ref{indens})).
Physically, this result is to be expected, since the growth of the
width of the probability distribution will be governed by the balance
between the bath fluctuations (\( S_{I} \)) and the friction (\( S_{R} \)).

\begin{figure}
{\centering \resizebox*{0.95\columnwidth}{!}{\includegraphics{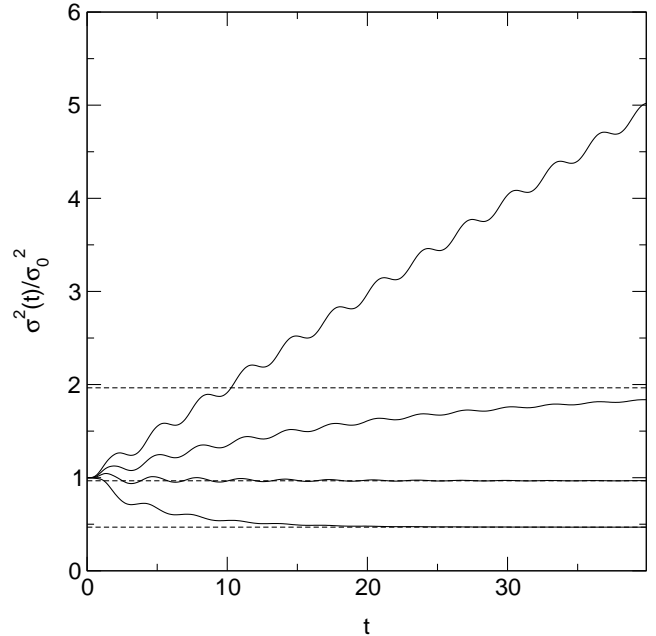}} \par}

\caption{\label{figeins}The variance \protect\( \sigma ^{2}(t)\equiv \left\langle \hat{q}^{2}(t)\right\rangle \protect \)
of the probability density \protect\( \rho (q,q,t)\protect \) for
the damped harmonic oscillator, plotted in units of the unperturbed
ground state variance \protect\( \sigma _{0}^{2}\protect \), as a
function of time \protect\( t\protect \) for the following friction
strengths (top to bottom): \protect\( \gamma /(2\pi \left\langle FF\right\rangle _{\omega =1})=0,\, 0.5,\, 1,\, 2\protect \)
(other parameters are \protect\( m=\omega _{0}=1\protect \)). A ratio
of \protect\( 1\protect \) corresponds to the correct value prescribed
by the FDT, while \protect\( \gamma =0\protect \) means \protect\( S_{R}\protect \)
has been neglected completely, such that \protect\( \sigma ^{2}(t)\protect \)
grows without bound. The dashed lines give the limit for \protect\( t\rightarrow \infty \protect \)
(see Eq.~(\ref{sigmalimit})).}
\end{figure}

The complete time-evolution of \( \left\langle \hat{q}^{2}(t)\right\rangle  \)
after switching on the interaction at \( t=0 \) is displayed in Fig.~\ref{figeins}.
This includes cases where the value of the damping constant \( \gamma  \)
has been artificially set to zero or to other values different from
that prescribed by the fluctuation-dissipation theorem (FDT), which
connects \( \gamma  \) to the strength of the bath fluctuations described
by \( \left\langle FF\right\rangle _{\omega } \) (see Eq. (\ref{ohmbathdef})
for the case of \( T=0 \)). If \( S_{R} \) is neglected completely
(no friction, \( \gamma =0 \)), heating takes place, causing the
variance to grow linearly in time, at a rate given by the Golden Rule
expression involving the \emph{symmetrized} bath correlator. Formally,
the growth of \( C_{00} \) with time cannot be compensated in that
case, since \( L_{00} \) and \( L_{t0} \) acquire their original
unperturbed values. For small \( \gamma >0 \), the width saturates
at a value much larger than that of the ground state. On the other
hand, if the friction is too strong, the width saturates at a value
below that of the ground state, and the Heisenberg uncertainty relation
is violated (since \( \left\langle \hat{p}^{2}\right\rangle  \) also
shrinks). This point is discussed in the book of Milonni\cite{milonni}
in the context of an atom interacting with the vacuum electromagnetic
field. If, however, \( \gamma  \) has the correct value prescribed
by the FDT, \( \left\langle \hat{q}^{2}\right\rangle  \) changes
only slightly from its unperturbed value.

In fact, for a linear system like the quantum harmonic oscillator,
its behaviour under the action of the linear bath can be described
and understood entirely using a classical picture\cite{weiss}. The
reason for this is as follows: one starts out with an initial state
for system and bath that has a positive definite Gaussian Wigner-density,
which can be interpreted directly as a classical phase space density
whose time-evolution then corresponds one-to-one to the evolution
of the Wigner-density (see appendix \ref{appWigner}). Thus, the quantum
dissipative dynamics may be described by a \emph{classical} Langevin
equation including friction and a fluctuating driving force:

\begin{equation}
\label{langevin}
\frac{d^{2}q}{dt^{2}}+\gamma \frac{dq}{dt}+\omega _{0}^{2}q=\frac{1}{m}F(t)\, .
\end{equation}

In this equation, the damping term incorporates the effects of \( S_{R} \).
It counteracts the fluctuations of \( F(\cdot ) \) that are described
by \( S_{I} \) in the path-integral picture and whose correlation
function is, therefore, given by the \emph{symmetrized} part of the
bath correlator.

The behaviour of the width has been discussed above using the influence
functional, but it can be understood more easily in this picture.
We can use the susceptibility of the damped oscillator to find its
response to the force fluctuations \( F \), which, at zero temperature,
are entirely due to the zero-point fluctuations. In particular, for
the limit \( t\rightarrow \infty  \), we obtain 

\begin{equation}
\label{sigmalimit}
\left\langle \hat{q}^{2}(t)\right\rangle \rightarrow \int _{-\infty }^{+\infty }d\omega \, \left| \chi (\omega )\right| ^{2}\left\langle FF\right\rangle _{\omega }\, ,
\end{equation}
where 

\begin{equation}
\label{susc}
\chi (\omega )=\frac{1/m}{\omega _{0}^{2}-\omega ^{2}-i\omega \gamma }
\end{equation}
is the susceptibility of the damped oscillator, and \( \left\langle FF\right\rangle _{\omega } \)
is the symmetrized power spectrum (see Eq. (\ref{ohmbathdef})). For
the correct equilibrium state (obtained by fully keeping \( S_{R} \)
and \( S_{I} \)), the width at \( t\rightarrow \infty  \) can also
be found by applying the FDT to the oscillator coordinate (cf. Eq.~6.37
of Ref.~\onlinecite{callegg}). In contrast, the effect of dropping
\( S_{R} \) is obtained by letting \( \gamma \rightarrow 0 \) in
Eq. (\ref{langevin}), while \( \left\langle FF\right\rangle _{\omega } \)
is kept fixed.

The example of the damped harmonic oscillator demonstrates that the
proper behaviour near the ground state of the system cannot be observed
at the early stages of the calculation, by looking at the action evaluated
along classical paths (Eqs.~(\ref{res}),~(\ref{ims})), regardless
of whether these paths properly include the damping or not. Only after
correct integration over the initial state density matrix the contributions
from \( S_{R} \) and \( S_{I} \) can compensate each other (in Eq.~(\ref{width})),
such that no artificial finite decay rate of the ground state results,
if \( S_{R} \) is kept.

\section{Power-law behaviour in quantum Brownian motion}

\label{qubrownmotion}We now turn to a discussion of the free particle,
i.e. the Caldeira-Leggett model of quantum Brownian motion. There,
the reasoning is the same as before in principle: Neglecting \( S_{R} \)
means that the zero-point fluctuations of the bath contained in \( S_{I} \)
may heat up the particle. However, there is an important difference:
The spectrum of the free motion does not contain any resonance peak,
unlike that of the harmonic oscillator. Therefore, only the low-frequency
components (\( \omega \rightarrow 0 \)) of the bath spectrum will
contribute to heating and decay (compare Eq.~(\ref{overlapspectra})),
which results in a peculiar long-time behaviour\cite{callegg,hakimambeg,schramm}
that cannot be captured using the Golden Rule, as has been emphasized
in Ref.~\onlinecite{GZ_CL}. 

One can observe the distinction between oscillator and free particle
most easily in the picture of the classical Langevin equation (\ref{langevin})
introduced above. In appendix \ref{appWigner}, it is shown how the
density matrix propagator \( J \) may be obtained in general via
the time-evolution of the Wigner density. This approach is physically
more transparent than the path-integral calculation. For the purposes
of our present discussion, however, it will suffice to consider the
time-evolution of the diagonal elements of the density matrix in momentum
space. These have also been analyzed in Ref.~\onlinecite{GZ_CL},
in order to demonstrate the failure of the Golden Rule calculation
to describe the decay of the original ground state at zero temperature.
We will confirm and explain the outcome of this analysis using a simple
argument based on the Langevin equation (\ref{langevin}). Then we
will point out the role of \( S_{R} \) and describe the important
difference to the damped oscillator.

If the particle has momentum \( p_{0} \) at time \( 0 \), its momentum
\( p \) at time \( t \) will be determined by the fluctuating force
\( F(\cdot ) \) in the following way (by solving Eq. (\ref{langevin})
for \( \omega _{0}=0 \)):

\begin{equation}
\label{pevol}
p=p_{0}e^{-\gamma t}+\int _{0}^{t}ds\, e^{-\gamma (t-s)}F(s)\, .
\end{equation}

Since \( F(\cdot ) \) is a Gaussian random process, the probability
density of finding a momentum \( p \) at time \( t \) is a Gaussian,
of variance

\begin{equation}
\label{pgaussian}
\left\langle \delta p^{2}\right\rangle \equiv \int _{0}^{t}ds_{1}\int _{0}^{t}ds_{2}e^{-\gamma (2t-s_{1}-s_{2})}\left\langle F(s_{1})F(s_{2})\right\rangle \, .
\end{equation}

The behaviour of \( \left\langle \delta p^{2}\right\rangle  \) as
a function of time is as follows (at \( T=0 \)): As long as \( t\ll 1/\gamma  \),
friction is unimportant and it is the double time-integral over the
correlator of \( F \) which is to be evaluated. For the Ohmic bath
considered here, the power spectrum is relatively strong at low frequencies
(rising linearly with \( \omega  \), see Eq. (\ref{ohmbathdef})).
This leads to a slow decay in time, \( \left\langle F(t)F(0)\right\rangle \propto 1/t^{2} \),
which results in a logarithmic growth of \( \left\langle \delta p^{2}\right\rangle  \),

\begin{equation}
\label{p2log}
\left\langle \delta p^{2}\right\rangle \approx 2\frac{\eta }{\pi }\ln (\omega _{c}t)\, .
\end{equation}
Here \( \omega _{c} \) is the cutoff frequency of the bath spectrum
and we have assumed \( \omega _{c}t\gg 1 \). It is this logarithmic
behaviour which cannot be obtained using the Golden Rule approximation,
to be discussed below. At later times, \( t\gg 1/\gamma  \), the
growth saturates at a constant value,

\begin{equation}
\left\langle \delta p^{2}\right\rangle \rightarrow \frac{\eta }{\pi }\ln (\omega _{c}/\gamma )\, .
\end{equation}

The perturbation expansion in the coupling between system and bath
can be carried out by expanding the influence action in the propagator
\( J \) (Eq. (\ref{prop})) in powers of \( iS_{R}-S_{I} \) (see
Eq. (\ref{eq13}) in the present article or Eq. (E13) of Ref. \onlinecite{GZ_CL}).
We want to discuss the time-evolution of the diagonal elements \( \rho _{pp}(t) \)
of the density matrix in momentum space. The evolution starts from
the equilibrium density matrix, which is a Gaussian momentum distribution
of variance \( \left\langle p_{0}^{2}\right\rangle =mT \) (the following
formulas will be analyzed for arbitrary \( T \)). After a time \( t \),
the density matrix is still a Gaussian, but of variance (compare Eq.
(\ref{pevol})):

\begin{equation}
\label{momvar}
\left\langle p^{2}\right\rangle =\left\langle p_{0}^{2}\right\rangle e^{-2\gamma t}+\left\langle \delta p^{2}\right\rangle \, .
\end{equation}

From this we can easily derive the result for \( \delta \rho ^{(1)}_{pp}(t) \)
which could alternatively be obtained by first expanding \( J \)
to first order in \( iS_{R}-S_{I} \), going over to the momentum
representation and finally integrating over the initial density matrix
(as was done in Ref.~\onlinecite{GZ_CL}; see Eq. (E18), which is,
however, still written in terms of matrix elements of the coordinate
operator):

\begin{eqnarray}
 &  & \rho _{pp}(t)=\frac{1}{\sqrt{2\pi \left\langle p^{2}\right\rangle }}\exp \left[ -\frac{p^{2}}{2\left\langle p^{2}\right\rangle }\right] \nonumber \\
 &  & \approx \rho _{pp}(0)\cdot (1+\frac{\left\langle p^{2}\right\rangle -\left\langle p_{0}^{2}\right\rangle }{2\left\langle p_{0}^{2}\right\rangle }\left( \frac{p^{2}}{\left\langle p_{0}^{2}\right\rangle }-1\right) +\cdots )\, .\label{rhoexpansion} 
\end{eqnarray}

In evaluating \( \left\langle p^{2}\right\rangle  \), only the terms
up to first order in \( \gamma  \) and \( \left\langle FF\right\rangle _{\omega } \)
must be kept for the purposes of this expansion. At finite temperature
\( T \), the behaviour of \( \left\langle p^{2}\right\rangle  \)
at long times \( t\gg 1/T \) is governed by the linear decrease stemming
from \( \left\langle p_{0}^{2}\right\rangle e^{-2\gamma t}\approx \left\langle p_{0}^{2}\right\rangle (1-2\gamma t) \)
and the linear increase from \( \left\langle \delta p^{2}\right\rangle \approx 2\eta Tt \).
If the initial density matrix really describes the equilibrium distribution,
then detailed balance holds and these terms cancel. For times \( t\ll 1/T \)
(or arbitrary times at \( T=0 \)), the time-evolution of \( \left\langle p^{2}\right\rangle  \)
is governed by \( \left\langle \delta p^{2}\right\rangle  \), evaluated
for the correlator \( \left\langle FF\right\rangle _{\omega } \)
taken at \( T=0 \) (which only contains the zero-point fluctuations
of the bath). For consistency of the expansion, \( \left\langle \delta p^{2}\right\rangle  \)
in Eq. (\ref{pgaussian}) has to be evaluated by setting \( \gamma =0 \).
Therefore, only the discussion given above for times \( t\ll 1/\gamma  \)
turns out to be relevant. The logarithmic growth of \( \left\langle \delta p^{2}\right\rangle  \)
given in Eq. (\ref{p2log}) corresponds to a power-law behaviour of
the (exact) time-evolution of \( \rho _{pp}(t) \) for \( t\ll 1/\gamma  \).
This logarithm, that appears in the perturbation expansion (\ref{rhoexpansion}),
is {}``overlooked'' in the Golden Rule approximation, where only
terms growing linearly with time are kept. Therefore, the Golden Rule
rate turns out to vanish\cite{GZ_CL}. These power-laws are characteristic
of the density matrix propagator \( J \) of quantum Brownian motion
at zero temperature (see appendix \ref{appWigner}).

On the other hand, the effect of neglecting \( S_{R} \) may be analyzed
by setting the friction constant \( \gamma  \) to zero in Eq. (\ref{pgaussian}).
Then, the spread \( \left\langle \delta p^{2}\right\rangle  \) of
the momentum distribution grows without bounds, which is obviously
not the correct physical behaviour. It is qualitatively similar to
the heating produced in the damped oscillator model. However, since
the growth proceeds only logarithmically in time, the artefacts of
this approximation, when applied to the free particle, are not nearly
as drastic as the finite decay rate of the ground state observed for
the oscillator. In particular, within perturbation theory, no qualitative
change is obtained by dropping \( S_{R} \) for the free particle
(no finite decay rate is produced). 

Furthermore, if we had considered a super-Ohmic bath (e.g. due to
phonons), whose spectrum decays faster than \( \omega ^{1} \) for
\( \omega \rightarrow 0 \), the growth of \( \left\langle p^{2}\right\rangle  \)
would saturate even without friction (as it should, since there is
no velocity-dependent friction for such a bath). In contrast, the
behaviour of the harmonic oscillator discussed above would remain
qualitatively the same as for the Ohmic bath, since its decay depends
on the bath spectrum at the resonance frequency of the oscillator
and is not affected in any essential way by the details of the spectrum
at low frequencies. This is consistent with the fact that in the oscillator
case already the Golden Rule is sufficient to obtain an essentially
correct picture of the artificial decay produced by dropping \( S_{R} \),
while, for quantum Brownian motion, it fails to describe the subtle
power-laws associated with the low-frequency (long-time) properties
of the Ohmic bath at zero temperature. 

The difference between oscillator and free particle also shows up
clearly in the behaviour of the imaginary part of the influence action:
For the free particle, \( S_{I} \) \emph{always} grows only logarithmically
with time \( t \) (at \( T=0 \)), independent of whether one keeps
\( S_{R} \) or sets it to zero (i.e. \( \gamma \equiv 0 \) in the
equations of motion). This can be derived from the formulas given
in appendix \ref{appWigner} or those of Ref.~\onlinecite{GZ_CL},
Appendix E (or Ref.~\onlinecite{callegg}, section 6, with the proper
limit \( \omega _{R}\rightarrow 0 \) for the free particle). 

Hence, we conclude that for the oscillator (or any motion containing
an extended spectrum) the effects of dropping \( S_{R} \) are much
more pronounced than for the free motion, since they lead to an artificial
finite decay rate. We have also observed that this effect can already
be understood within the framework of the Golden Rule approximation,
whereas that approximation is incapable of describing the more subtle
{}``power-law decay'' found for the free particle.

\section{Qualitative discussion of other models: Nonlinear coupling and Pauli
principle}

\label{section7}It could be argued that our discussion of the damped
oscillator has only demonstrated the importance of \( S_{R} \) in
preventing an artificial decay of the \emph{ground state} and has
little to do with dephasing. Indeed, if one wants to discuss dephasing,
one should rather consider the decay of a coherent superposition of
the ground state and some excited state. This is relevant both for
arbitrary nonequilibrium situations as well as for the calculation
of the system's linear response, where the perturbation creates a
superposition of excited states and the ground state. In the example
of the damped harmonic oscillator, full relaxation into the ground
state will \emph{always} take place (at \( T=0 \)), for any such
excited state. This holds regardless of the bath spectrum, provided
the latter does not vanish around the resonance frequency of the oscillator.
For weak coupling, the decay of the density matrix (and, therefore,
the linear response) may be described conveniently using a simple
master equation approach. What is more, even if one neglects \( S_{R} \)
in a path-integral calculation, the predicted decay rates will be
correct, excepting only that for the ground state, which we have discussed
above. The reason for this, however, is a specialty of the linearly
damped harmonic oscillator. Since the bath couples to the coordinate
operator, whose matrix elements only connect adjacent oscillator levels,
the relevant {}``system correlator'' (in Eq.~(\ref{systemspec}))
turns out to be symmetric for any state but the lowest one. Therefore,
the decay rate will depend only on the symmetric part of the bath
correlator as well, see Eq. (\ref{pathmaster}). Hence, for all the
excited states, the evaluation of the total decay rate is not affected
by dropping \( S_{R} \). This point has already been discussed in
connection with the model of two oscillating wave packets, at the
end of section \ref{section2}.

\begin{figure}
{\centering \resizebox*{0.95\columnwidth}{!}{\includegraphics{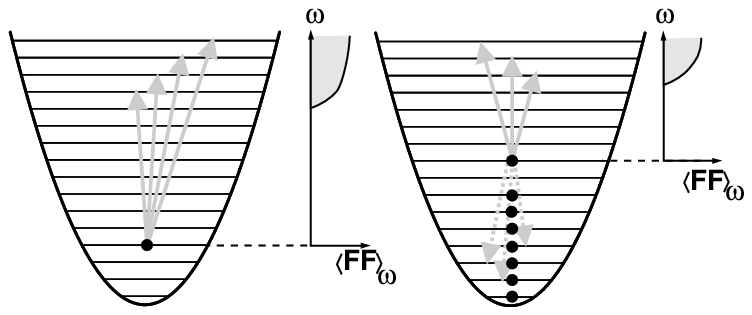}} \par}

\caption{\label{fig2}The models described in the text. \emph{Left}: A bath
with a gap would lead to the upward transitions indicated by grey
arrows if (and only if) \protect\( S_{R}\protect \) were neglected.
\emph{Right}: A sea of fermions may block downward transitions (dashed)
even though they are possible in the single-particle picture. }
\end{figure}

Since we are interested in demonstrating the importance of \( S_{R} \)
for the dephasing of a superposition of low-lying levels, the natural
choice is to consider a bath with an excitation gap larger than the
resonance frequency of the oscillator (see Fig.~\ref{fig2}). In
that case, we expect the lowest levels to remain coherent in the correct
description. Unfortunately, in the model of the linearly damped oscillator
discussed above, we would not obtain any finite decay rate, \emph{regardless}
of whether \( S_{R} \) is kept or not. This is because the corresponding
classical noise force is out of resonance and cannot heat the oscillator. 

Therefore, in order to analyze a situation where the importance of
retaining \( S_{R} \) is displayed even for excited states, we have
to go beyond exactly solvable linear models. As a consequence, the
following discussion is necessarily incomplete insofar as we cannot
give exact proofs of the statements to be made below. These statements
will be based on the experience acquired in the simpler models discussed
above. We will also make use of some Golden Rule type arguments, noting
that the Golden Rule has been sufficient to understand the behaviour
of the damped oscillator in a qualitatively correct way. In any case,
we believe it is useful to contrast the results obtained by dropping
\( S_{R} \) with the {}``commonly accepted'' picture.

Let us, therefore, consider the following model: We retain the harmonic
oscillator, but change the coupling between system and bath from \( \hat{V}=-\hat{q}\hat{F} \)
to \( \hat{V}=-f(\hat{q})\hat{F} \), with a \emph{nonlinear} function
\( f \). This is called state-dependent friction\cite{weiss}. In
that case, the transition matrix elements of the system operator \( f(\hat{q}) \)
will, in general, be nonzero between any two states, which is the
important difference to the linear case. If we introduce a bath spectrum
with a gap larger than \( n\omega _{0} \), then, for sufficiently
weak coupling and according to a simple Golden Rule calculation, the
first \( n \) excited states will not decay at zero temperature.
Any coherent superposition of these states will therefore remain coherent.
The full, non-perturbative picture will be slightly more complicated,
but the essential point should remain the same: The first excited
states will acquire an admixture of other levels and become entangled
with the states of the bath. They will experience some frequency shifts
and the transition matrix elements will be renormalized. Of course,
the new eigenstates of the coupled system will remain in a coherent
superposition forever (by definition). The \( n+1 \) lowest ones
still have a discrete spectrum and are in direct correspondence to
the initial unperturbed states. When switching on the interaction
at \( t=0 \), a partial decay will result. Physically, this corresponds
to the relaxation of the initial state into the selfconsistent coupled
state of system and bath (see Ref.~\onlinecite{hakimambeg} for a
discussion of these issues in the case of the free particle). However,
the initial decay will saturate in time on a short time-scale. During
this transient adjustment, the reduced system density matrix becomes
mixed to a small extent (if the coupling is weak), but no non-saturated
long-term decay (indicated by a finite decay rate) results. In any
case, the transient decay is an artefact of the procedure of suddenly
switching on the interaction. A full calculation of, for example,
the linear response properties of the dissipative system would start
out with the selfconsistent ground state of the fully interacting
system (as was done in Ref.~\onlinecite{schramm} for the free particle).

In contrast, dropping \( S_{R} \) would lead to a completely different
picture, since then \( S_{I} \) would correspond to a classical noise
force. Starting from any of the first \( n \) excited states, the
fluctuating field would always be able to induce transitions \emph{up}
in energy, thus leading to a finite decay rate. The role of \( S_{R} \)
in the correct description of a quantum noise force is precisely to
cancel these upward transitions, as has been demonstrated in the case
of the linearly damped oscillator. Due to this finite decay rate,
any initially coherent superposition would get destroyed, leading
to complete dephasing. No saturation of the decay could be observed,
in spite of the gap in the bath spectrum. 

It is \emph{this} behaviour that is described qualitatively correctly
by evaluating \( S_{I} \) along semiclassical paths of the oscillator.
The expression for \( S_{I} \), Eq.~(\ref{silin}), is now to be
changed simply by replacing \( q \) by \( f(q) \). While an (unperturbed)
oscillator path of the form \( q(\tau )\propto \cos (\omega _{0}t) \)
only contains frequencies \( \pm \omega _{0} \) which cannot couple
to the gapped bath, the function \( f(q(\tau )) \) will, in general,
contain all higher harmonics as well. This directly corresponds to
the character of the \emph{symmetrized} system spectrum, i.e. the
Fourier transform of \( \left\langle \left\{ f(\hat{q}(\tau )),f(\hat{q}(0))\right\} \right\rangle  \).
These frequencies contained in the system motion then will couple
to the bath spectrum, leading to an unbounded growth of \( S_{I} \)
with time. 

Similarly, if we were to use a bath without a gap, but where the spectrum
falls off fast towards low frequencies, then the correct description
would also yield decay rates that quickly become smaller when going
towards the ground state, while they would saturate at a finite, comparatively
large level if \( S_{R} \) were dropped. 

The present work has been motivated in part by the ongoing controversial
debate on low-temperature dephasing in an interacting disordered fermion
system\cite{mohantywebb,GZ,GZ_CL,GZ_PB,cohenimry,VavAmbeg,AAGcritique,BelitzKirckpatrick},
which has led to a revival of interest for the general question of
dephasing in mesoscopic systems\cite{CedrBuett,21,seelig,buettHO,Guinea,ghz}.
Since the usual Feynman-Vernon influence functional deals only with
a single-particle situation, it cannot be applied directly to this
problem. However, the authors of Refs.~\onlinecite{GZ,GZ_CL,GZ_PB}
have succeeded in deriving an extension of the usual influence functional
to the many-fermion situation, using an exact procedure. In their
result, the Fermi distribution (and, therefore, the Pauli principle)
only enters \( S_{R} \), while \( S_{I} \) is unaffected (Eqs. (54)
and (55) of Ref.~\onlinecite{GZ}). Therefore, the dephasing rate,
as read off from \( S_{I} \), is found to equal the rate which would
also be obtained in a purely single-particle calculation. Although
we do not analyze the full problem of weak-localization here, we will
use the insights gained above in order to explain why, in our opinion,
the \emph{assumption} that the dephasing rate can be derived from
\( S_{I} \) alone must be proven instead of being taken for granted.
This holds even when \( S_{R} \) vanishes on the relevant pairs of
classical time-reversed trajectories, since, in general, the integration
over the fluctuations away from these trajectories will be essential
for obtaining a cancellation between the effects of \( S_{R} \) and
\( S_{I} \) (see sec. \ref{section4}), in addition to properly taking
into account the initial density matrix. Furthermore, we note that
in these calculations the detailed form of the bath spectrum at low
frequencies turns out to be unimportant for the essential result of
a finite zero-temperature dephasing rate. This is in marked contrast
to the case of the free particle (where the Ohmic bath plays a special
role\cite{schramm}; see also Refs. \onlinecite{Guinea,ghz}), but
similar to what is observed for the damped harmonic oscillator when
\( S_{R} \) is neglected. It is for this reason that we have chosen
a bath with an excitation gap in order to demonstrate the importance
of \( S_{R} \), since there the effects come out most clearly, although
they are also present for other bath spectra (compare the discussion
above).

The model situation of the damped oscillator described above already
contains one key ingredient related to the description of dephasing
for an electron moving inside a disordered metal, namely an extended
system spectrum: Due to the impurity scattering, the system spectrum
(e.g., of the velocity operator) contains frequencies up to (at least)
the elastic scattering rate, which may couple to the zero-point fluctuations
contained in the bath spectrum. This is in contrast to the free particle,
which can only couple to the low-frequency bath modes (see the discussion
at the end of section \ref{qubrownmotion}). As before, the importance
of \( S_{R} \) will be visible most easily for a bath containing
an excitation gap. For such a bath, the \emph{free} motion (without
impurity scattering) will not show any nontransient decay, \emph{regardless}
of whether \( S_{R} \) is taken into account or not. In contrast,
the model to be discussed below, which is more similar to the situation
in the disordered metal, will show a finite decay rate of low-lying
levels if and only if \( S_{R} \) is neglected.

Apart from the extended system spectrum, we have to take into account
another important feature of the calculations\cite{GZ,GZ_CL,GZ_PB,cohenimry,AAK}
concerning electrons in a disordered metal: the use of the semiclassical
analysis. For linear systems the semiclassical result for the path-integral
is exact, such that it can even be used near the ground state of the
system, as we have done it here. However, in such a case, the size
of the fluctuations around the path is comparable to the amplitude
of the path itself, which is not the situation in which semiclassics
is usually applicable. In order to gain intuition for a typical semiclassical
situation, we could reconsider the example with the nonlinear coupling
to a gapped bath given above: Imagine an initial superposition of
fast wavepackets, whose wavelength is much smaller than the packet
size but whose excitation energies are still below the bath threshold.
They would show a finite decay or dephasing rate (for the reasons
given above), if only \( S_{I} \) were used for the analysis, but
not in the full calculation. Still, for this scenario one might argue
that it has been clear from the outset that the semiclassical analysis
cannot be trusted, since those high-frequency components in the system
spectrum that are responsible for the decay are necessarily larger
than the energetic distance to the ground state. 

It is only in a system of degenerate fermions that the following two
conditions can be fulfilled all at once: On the one hand, the semiclassical
analysis of a single (non-interacting) electron moving at the Fermi
level is valid for an external potential which is sufficiently slowly
varying, such that the system spectrum (concerning the motion of the
single electron) only contains frequencies much smaller than the Fermi
energy. On the other hand, the whole many-particle system may be near
(or in) its ground state. 

While the first feature would lead one to believe that \( S_{R} \)
may be omitted, we have learned from the examples discussed before
that this is likely to be incorrect whenever the second condition
holds as well. We stress once again that the two conditions are mutually
exclusive in a single-particle problem, which is the reason why such
considerations have not played any role in influence-functional calculations
up to now. 

In order to render the discussion concrete, we can once again make
use of the example with a nonlinear coupling \( f(\hat{q}) \) given
above, provided we suppose the \( N \) lowest oscillator states to
be filled up with fermions initially (see Fig.~\ref{fig2}). For
a relatively smooth function \( f \), the matrix elements of \( f(\hat{q}) \)
only connect states within a range much smaller than \( \epsilon _{F}=N\omega _{0} \),
which, in our example, should still be larger than the gap \( n\omega _{0} \)
of the bath. Dephasing of a \emph{single} particle near \( \epsilon _{F} \)
can then certainly be described fully within the semiclassical analysis,
and, to a good approximation, using \( S_{I} \) alone, as has been
discussed above. The same holds for the many-particle system, if one
explicitly considers a \emph{classical} noise force, where \( S_{R} \)
is absent. In that case, the many-particle problem can really be treated
as a collection of independent single-particle problems, as is the
case for any external time-dependent potential. Only in the end an
average of the full Slater determinant over all possible realizations
of the external noise has to be carried out. 

If the dephasing rate is calculated solely from \( S_{I} \), it turns
out to be finite\cite{GZ,GZ_CL,GZ_PB} and not to depend at all on
the distance to the Fermi surface. This is consistent with the fact
that the the value of \( \epsilon _{F} \) does not even appear in
the single-particle calculation. On the other hand, for a quantum
bath \( S_{R} \) does not vanish and should be included in the influence
action. It is true that for the case of a highly excited \emph{single}
particle the dephasing rate comes out correct, regardless of whether
\( S_{R} \) is kept or not, as we have discussed before. However,
we have also pointed out that there is an important physical difference
between the two calculations. Since in the correct approach (including
\( S_{R} \)), the transitions induced by the bath are purely downwards
in energy (accompanied by spontaneous emission into the bath), it
is reasonable to expect that, in the \emph{many-particle} problem,
they will be blocked by the Pauli principle. In any case, this is
what is found using the Golden Rule. Judging from these arguments,
the contribution to \( S_{I} \) from the nonzero overlap of the (symmetrized)
system spectrum with the high frequencies of the bath spectrum does
not imply dephasing and decay, but rather a renormalization, similar
to that obtained for an electron interacting with optical phonon modes
(a gapped bath), leading to the formation of a polaron. The formation
of the polaron will be visible as an initial transient decay of the
single-electron density matrix (for the artificial case of factorized
initial conditions), which saturates on a short time-scale. Therefore,
to lowest order, for a bath with a gap no finite relaxation rate of
the lowest-lying single-particle excitations above the Fermi sea is
expected, just as for the lower levels of the harmonic oscillator
in the single-particle model with nonlinear coupling to a gapped bath. 

We have to qualify this statement by taking into account the fact
that the coupling between electrons and bath always also induces an
effective interaction between the electrons. In this way, a given
electron becomes coupled indirectly to the bath of other electrons,
such that scattering processes will, indeed, lead to a finite decay
rate even for those low-lying excited levels. This precludes any rigorous
proof demonstrating the complete absence of decay and dephasing for
low-lying levels even in such a rather simple situation, in spite
of the assumption of an excitation gap in the bath spectrum. However,
the consequences of the effective interaction can be distinguished
easily from the finite dephasing rate that would be predicted by looking
at \( S_{I} \) alone. If the coupling to the bath is of strength
\( g \), then the latter rate would go as \( g^{2} \). In contrast,
the effective interaction (obtained after integrating out the bath)
will itself be of strength \( g^{2} \), such that the resulting relaxation
rate (due to coupling of an electron to the density fluctuations of
other electrons) will be of fourth order in \( g \). In addition,
of course, the rate will depend strongly on the distance to the Fermi
surface, vanishing when the Fermi energy is approached.

\section{Conclusions}

We have tried to demonstrate that the real part \( S_{R} \) of the
Feynman-Vernon influence action cannot be neglected in an analysis
of dephasing and decay \emph{near the ground state of a system}, in
spite of the fact that, in the simplest situations (involving highly
excited states and the semiclassical analysis), it is only the imaginary
part \( S_{I} \) which enters the dephasing rate. To this end, we
have discussed how \( S_{R} \) and \( S_{I} \) may cancel each other's
effects not only in lowest-order perturbation theory but also to all
orders, by examining exactly solvable linear quantum dissipative systems.
In general, the cancellation is only found after proper integration
over the fluctuations away from semiclassical trajectories, taking
into account the action of the unperturbed system and its initial
density matrix. Furthermore, we have pointed out an essential difference
between the damped oscillator and the free particle with respect to
this issue. We have argued that the insight obtained in the case of
the damped oscillator is also applicable to nonlinear systems and
important for discussions of dephasing in systems of disordered degenerate
fermions.

In summary, it may be possible to discuss dephasing and decay without
considering \( S_{R} \) either if the noise is nearly classical (external
nonequilibrium radiation or bath in the high-temperature limit) or
if the system itself is in a highly excited state. Otherwise, dephasing
rates obtained solely from \( S_{I} \) are bound to come out finite
at zero temperature in most cases, even when a simpler Golden Rule
calculation gives vanishing results. Judging from the examples discussed
above, this is probably not because such a nonperturbative procedure
goes beyond the Golden Rule approximation, but because it neglects
some physics already contained even within this approximation. 

Still, although the primary message of this paper is that special
care has to be taken in extracting dephasing rates from calculations
using the influence functional, we emphasize at the same time that
there is no \emph{general} proof showing the impossibility of zero-temperature
dephasing near the ground state. There cannot be such a proof, since
there is evidently at least one counter-example (the Caldeira-Leggett
model of quantum Brownian motion and similar models\cite{callegg,Guinea,ghz}),
and, furthermore, there is no generally applicable definition of {}``dephasing''
that is useful under all conceivable circumstances. 

\begin{acknowledgments}I would like to thank Christoph Bruder, Jan
von Delft, Dimitri Golubev and Andrei Zaikin for stimulating discussions.
This work has been funded by the Swiss National Science Foundation.\end{acknowledgments}

\appendix

\section{Some quantities for the damped oscillator}

\label{appc00}In this appendix, we list, for purposes of reference,
the quantities relevant to our discussion of the time-evolution of
the quantum damped harmonic oscillator. These can also be found in
Ref. \onlinecite{callegg} (in a slightly different notation).

The quantity \( C_{00} \) arises in evaluating \( S_{I} \) along
a pair of semiclassical paths, by inserting the solution \( r(\cdot ) \)
of Eq. (\ref{req}) for the boundary conditions given by \( r_{t},\, r_{0} \)
and determining the coefficient of \( r_{0}^{2} \) in the result
(see Eq. (\ref{ims})). Necessarily, \( C_{00} \) contains only the
symmetric part of the bath correlator, since \( S_{I} \) depends
only on that: 

\begin{eqnarray}
C_{00}=\frac{1}{4}\left( \sin (\tilde{\omega }t)\right) ^{-2}\int _{0}^{t}dt_{1}\int _{0}^{t}dt_{2}\, \exp \left( \frac{\gamma }{2}(t_{1}+t_{2})\right)  &  & \nonumber \\
\times \sin (\tilde{\omega }(t-t_{1}))\left\langle \{\hat{F}(t_{1}),\hat{F}(t_{2})\}\right\rangle \sin (\tilde{\omega }(t-t_{2}))\, . &  & \label{C00formula} 
\end{eqnarray}

Here \( \tilde{\omega }\equiv \sqrt{\omega _{0}^{2}-(\gamma /2)^{2}} \)
is the renormalized frequency of the (underdamped, \( \gamma <2\omega _{0} \))
oscillator. 

The other quantities needed for calculating \( \left\langle \hat{q}^{2}(t)\right\rangle  \)
arise from the evaluation of \( Re\, S_{cl} \) (Eq. (\ref{res})).
In contrast to \( C_{00} \), which has to be evaluated numerically,
they can be given in closed form:

\begin{eqnarray}
L_{00} & = & m(\gamma /2+\tilde{\omega }\cot (\tilde{\omega }t))\\
L_{t0} & = & -\frac{m\tilde{\omega }e^{\gamma t/2}}{\sin (\tilde{\omega }t)}
\end{eqnarray}

The exponential increase of \( L_{t0} \) cancels that of \( C_{00} \)
when calculating the width in Eq. (\ref{width}). However, if the
damping rate \( \gamma  \) is set to zero, \( C_{00} \) still grows
beyond all bounds while \( L_{t0} \) remains bounded.

\section{Density matrix propagator from the Wigner density evolution}

\label{appWigner}Using the classical Langevin equation (\ref{langevin}),
the kernel \( J \) which relates the reduced density matrix of a
linear damped system at time \( t \) to that at time \( 0 \) can
be found in a way which is physically more transparent than the corresponding
derivation using path-integrals (see Eqs. (\ref{rhoJ}), (\ref{prop}),
(\ref{Jsemi}), (\ref{res}) and (\ref{ims})). One first solves for
the time-evolution of the classical phase space density (i.e. the
Wigner density) under the action of friction and the Gaussian random
force \( F(\cdot ) \). Starting from a \( \delta  \) peak located
in phase space, the phase space density evolves into a two-dimensional
Gaussian distribution, whose covariances are related to the correlator
of the force. This gives the propagator \( J^{W} \) of the Wigner
density, which only needs to be Fourier transformed with respect to
the momenta in order to obtain the density matrix propagator \( J \),
expressed via center-of-mass and difference coordinates \( R \) and
\( r \):

\begin{eqnarray}
J(R_{t},r_{t}|R_{0},r_{0};t)= &  & \nonumber \\
\int \, dp_{t}dp_{0}\, e^{i(p_{t}r_{t}-p_{0}r_{0})}J^{W}(R_{t},p_{t}|R_{0},p_{0};t)\, . &  & \label{propWigner} 
\end{eqnarray}

Here we show how the propagator \( J \) for the density matrix of
a free damped particle subject to the Ohmic bath may be obtained in
this way. Everything works the same for the damped oscillator, only
the resulting expressions are slightly more lengthy.

The propagator \( J^{W}(R_{t},p_{t}|R_{0},p_{0};t) \) of the Wigner
density is found by solving the classical equations of motion for
\( R \) and \( p \) for a given initial condition \( (R_{0},p_{0} \)),
taking into account friction and the action of the force \( F \):

\begin{eqnarray}
\frac{dp}{dt} & = & -\gamma p+F(t)\\
\frac{dR}{dt} & = & \frac{p}{m}
\end{eqnarray}

This yields the solutions

\begin{eqnarray}
p_{t} & = & p_{0}e^{-\gamma t}+\xi _{p}\label{Wigpev} \\
R_{t} & = & R_{0}+\frac{p_{0}}{\eta }(1-e^{-\gamma t})+\xi _{R}\, ,\label{WigRevol} 
\end{eqnarray}

where \( \xi _{p} \) and \( \xi _{R} \) are given as integrals over
the force \( F(\cdot ) \):

\begin{eqnarray}
\xi _{p} & = & \int _{0}^{t}ds\, e^{-\gamma (t-s)}F(s)\label{xip} \\
\xi _{R} & = & \frac{1}{\eta }\int _{0}^{t}ds\, (1-e^{-\gamma (t-s)})F(s)\, .\label{xir} 
\end{eqnarray}
 Since \( F(\cdot ) \) is a Gaussian random process, \( \xi _{p} \)
and \( \xi _{R} \) are Gaussian random variables as well. Therefore,
the phase space density evolving out of \( \delta (R-R_{0})\delta (p-p_{0}) \)
is a two-dimensional Gaussian distribution in phase space \( (R,p) \):

\begin{equation}
J^{W}(R_{t},p_{t}|R_{0},p_{0};t)=\left\langle \delta (R_{t}-\bar{R}_{t}-\xi _{R})\delta (p_{t}-\bar{p}_{t}-\xi _{p})\right\rangle \, .
\end{equation}

The average values \( \bar{p}_{t} \) and \( \bar{R}_{t} \) may be
read off from eqs. (\ref{Wigpev}) and (\ref{WigRevol}). \( J^{W} \)
has to be Fourier transformed with respect to \( p_{t} \) and \( p_{0} \)
in order to arrive at the density matrix propagator \( J(R_{t},r_{t}|R_{0},r_{0};t) \)
(see Eq. (\ref{propWigner})). In order to do this, we express \( J^{W} \)
as a Gaussian density in terms of \( p_{t},\, p_{0} \), for fixed
\( R_{t},R_{0} \):

\begin{equation}
J^{W}\propto \exp \left[ -\frac{1}{2}\delta P^{t}K^{-1}\delta P\right] \, ,
\end{equation}

with

\begin{equation}
\delta P=\left[ \begin{array}{c}
p_{t}-\bar{p}_{t}\\
p_{0}-\bar{p}_{0}
\end{array}\right] 
\end{equation}

and the covariance matrix 

\begin{equation}
K=\left[ \begin{array}{cc}
\left\langle \delta p_{t}^{2}\right\rangle  & \left\langle \delta p_{t}\delta p_{0}\right\rangle \\
\left\langle \delta p_{t}\delta p_{0}\right\rangle  & \left\langle \delta p_{0}^{2}\right\rangle 
\end{array}\right] \, .
\end{equation}

Making use of (\ref{Wigpev}) and (\ref{WigRevol}), the average values
and the deviations of \( p_{t} \) and \( p_{0} \) are found to be
given by:

\begin{eqnarray}
\bar{p}_{t} & = & \lambda (R_{t}-R_{0})\\
\bar{p}_{0} & = & e^{\gamma t}\lambda (R_{t}-R_{0})\\
\delta p_{t} & = & \xi _{p}-\lambda \xi _{R}\label{dpt} \\
\delta p_{0} & = & -e^{\gamma t}\lambda \xi _{R}\label{dp0} \\
\lambda  & \equiv  & \frac{\eta }{e^{\gamma t}-1}
\end{eqnarray}

\( \bar{p}_{0} \) and \( \bar{p}_{t} \) are the momenta at time
\( t \) and \( 0 \) which the particle must have if it is to go
from \( R_{0} \) to \( R_{t} \) in time \( t \), provided no fluctuating
force acts. \( \delta p_{t} \) and \( \delta p_{0} \) are the deviations
from these values needed to compensate the effects of \( F(\cdot ) \). 

Quadratic completion in the exponent immediately yields the result
of the Fourier integration over \( p_{t} \) and \( p_{0} \), which
is the desired density matrix propagator \( J \):

\begin{eqnarray}
J(R_{t},r_{t}|R_{0},r_{0};t)\propto  &  & \nonumber \\
\exp \left[ i(R_{t}-R_{0})\lambda (r_{t}-e^{\gamma t}r_{0})-\frac{1}{2}\left\langle \left( r_{t}\delta p_{t}-r_{0}\delta p_{0}\right) ^{2}\right\rangle \right] \, . &  & \label{jprop} 
\end{eqnarray}

This reproduces the result given in Ref.~\onlinecite{callegg} (or
Ref.~\onlinecite{GZ_CL}, App. E). The prefactor can be determined
from the normalization condition, Eq. (\ref{normal}), and only depends
on the time \( t \). The term in angular brackets still has to be
averaged over the force \( F(\cdot ) \). The resulting real part
of the exponent equals \( -S_{I} \) evaluated along the semiclassical
paths (compare the general structure given in Eq. (\ref{ims})). In
terms of \( F(\cdot ) \), \( \delta p_{t} \) and \( \delta p_{0} \)
read explicitly:

\begin{eqnarray}
\delta p_{t} & = & \int _{0}^{t}ds\, \frac{e^{\gamma s}-1}{e^{\gamma t}-1}F(s)\\
\delta p_{0} & = & \int _{0}^{t}ds\, \frac{e^{\gamma s}-e^{\gamma t}}{e^{\gamma t}-1}F(s)\, .
\end{eqnarray}

Note that

\begin{equation}
\delta p_{t}-\delta p_{0}=\int _{0}^{t}ds\, F(s)\, .
\end{equation}

The averages \( \left\langle \delta p_{t}^{2}\right\rangle  \), \( \left\langle \delta p_{0}^{2}\right\rangle  \)
and \( \left\langle \delta p_{t}\delta p_{0}\right\rangle  \) to
be evaluated in Eq. (\ref{jprop}) contain time-integrals over the
force correlator \( \left\langle F(s_{1})F(s_{2})\right\rangle  \).
At zero temperature, these integrals lead to terms growing logarithmically
in time, which are characteristic for the free particle coupled to
an Ohmic bath (compare the discussion in the main text, sec. \ref{qubrownmotion}).


\begin{thebibliography}{10}
\bibitem{5}Note that the damping constant \( \gamma  \) used here corresponds
to \( 2\gamma  \) in Ref.~\onlinecite{callegg}, starting from their
Eq.~(6.8).
\end{thebibliography}
\end{document}